\documentclass[aps,prb,twocolumn,superscriptaddress,showpacs]{revtex4}
\usepackage{graphicx}
\usepackage{dcolumn}

\begin{document}

\title{Vibrations of free and embedded 
anisotropic elastic spheres: \\
Application to low-frequency Raman scattering 
of silicon nanoparticles in silica}

\author{Lucien Saviot}
\affiliation{Laboratoire de Recherche sur la R\'eactivit\'e des Solides,
UMR 5613 CNRS - Universit\'e de Bourgogne\\
9 avenue A. Savary, BP 47870 - 21078 Dijon - France}
\email{Lucien.Saviot@u-bourgogne.fr}

\author{Daniel B. Murray}
\affiliation{Department of Physics,
Okanagan University College,
Kelowna, British Columbia, Canada V1V 1V7}
\email{dmurray@ouc.bc.ca}

\author{Maria del Carmen Marco de Lucas}
\affiliation{Laboratoire de Recherche sur la R\'eactivit\'e des Solides,
UMR 5613 CNRS - Universit\'e de Bourgogne\\
9 avenue A. Savary, BP 47870 - 21078 Dijon - France}

\date{\today}

\begin{abstract}
Vibrational mode frequencies and damping are
calculated for an elastic sphere embedded in 
an infinite, homogeneous, isotropic elastic 
medium. Anisotropic elasticity of the sphere 
significantly shifts the frequencies in 
comparison to simplified calculations that 
assume isotropy.  New low frequency
Raman light scattering data are presented for
silicon spheres grown in a SiO$_2$ glass matrix.  
Principal features of the Raman spectrum are 
not correctly described by a simple model of 
the nanoparticle as a free, isotropic sphere, 
but require both matrix effects and the anisotropy
of the silicon to be taken into account.
Libration, not vibration, is the dominant mechanism.
\end{abstract}

\pacs{63.22.+m,78.30.-j,43.20.+g,02.70.Ns}
\maketitle

\section{Introduction}
The classical continuum mechanical problem of the
vibrational frequencies of a homogeneous, isotropic,
free sphere was solved exactly long ago in terms of
roots of spherical Bessel functions \cite{lamb1882}.
It finds application in seismology \cite{sato62},
foam spheres \cite{mcdaniel00}
and even kidney stones \cite{mitri03}.
Progress has also been made on
generalizations of this problem.
Closed form solutions are obtainable
for special cases of inhomogeneity with 
certain special kinds of anisotropy \cite{Chen00},
and isotropy with radial inhomogeneity \cite{PortalesPRB02}.  
Generalized anisotropy can be handled for
a sphere and other shapes with a
basis function expansion \cite{Visscher91}.

Application of the Lamb solution to metal or
semiconductor spheres a few nanometers in size has
been successful in
explaining vibrational frequencies observable
using low frequency Raman light scattering \cite{Duval86}
and other methods \cite{Voisin02}. These "nanoparticles" are
not truly free, but are embedded inside a
glass matrix and frequently are nearly
spherical. The Lamb solution is useful because
of the sharp change of acoustic impedance
$z = \rho c$ at the nanoparticle-glass
boundary as long as the nanoparticle is much
heavier or much harder than the glass. The
applicability of this is challenged in the
case of silicon nanoparticles grown in
SiO$_2$ glass, where the densities are almost
the same and the speed of sound in silicon is
only 1.5 times faster than in glass.
Also, silicon is far from isotropic
with Zener anisotropy factor 1.6.

The nanoparticle's vibrational frequencies 
become complex valued due to damping of the 
modes as energy is mechanically radiated away 
from the nanoparticle into the surrounding 
glass matrix \cite{dubrovskiy81}. 
The effect of coupling the nanoparticle
to the matrix is to shift the 
frequencies of the modes and dampen them \cite{Tamura82}.
Some qualitatively new vibrational modes 
appear even in the limit of weak coupling \cite{Ovsyuk96}.
The solution to the elastic mechanical problem 
of an elastically anisotropic sphere embedded 
in an infinite elastic continuum has not been 
presented before.  Here we employ both a closed
form solution for the case of isotropic 
elasticity as well as molecular dynamics (MD)
simulations for the case of a cubic crystal
with elastic constants $C_{11}$, $C_{12}$ 
and $C_{44}$.
The agreement of these two methods for the exactly
solved free sphere case allow us to be 
confident in the correctness of our solution.
Two different computer programs were written
for the embedded isotropic sphere calculation.
These cross checks are essential to avoid
errors that have occurred elsewhere in the literature
of this subject.

\section{Theory}
The displacement field 
$\vec u (\vec r,t)$ of an elastic medium of 
density $\rho(\vec r)$ is governed by
Navier's equation: (\ref{eq:anisotropic}).
\begin{equation}
  c_{ijkl,j} u_{k,l} + c_{ijkl} u_{k,lj}  = \rho \ddot{u}_i
  \label{eq:anisotropic}
\end{equation}
where $c_{ijkl}(\vec r)$ is the 4th rank 
elastic tensor field with 21 independent 
components.  In a homogeneous, isotropic 
medium, Eq.~\ref{eq:anisotropic} takes the form 
of Eq.~\ref{eq:isotropic} where $\lambda$ and
$\mu$ are Lam\'e's constants.
\begin{equation}
  \left( \lambda + 2 \mu \right) \vec \nabla         
        (\vec \nabla \cdot \vec u)
  - \mu \vec \nabla \times \left(\vec \nabla \times \vec u \right)
  = \rho \ddot{ \vec u }
  \label{eq:isotropic}
\end{equation}

To elucidate the key features and physical basis of the motion,
the vibrations of the nanoparticle are calculated here
using three methods: (A) analytic solution of a free
isotropic sphere reprising reference \cite{lamb1882};
(B) analytic solution of an isotropic sphere embedded in
an infinite matrix; (C) MD computer simulation of a
free sphere including anisotropy.
Method (B) has previously been used only
for some spheroidal modes \cite{Voisin02,Verma99}
or limiting cases \cite{dubrovskiy81}. Method (C) is
reported for the first time.

The following comments apply to methods (A) and (B)
which are able to exploit spherical symmetry. The
time dependence of the displacement is taken as 
$  \vec u ( \vec r , t ) = \vec u ( \vec r ) e^{i \omega t}, $
where $\omega$ can be complex.
The angular frequency $\omega$ (in rad/s) is related to
wavenumber $\nu$ (in cm$^{-1}$) through $ \nu = \omega / ( 200 \pi c ) $,
where $c$ is the speed of light (in m/s).  In what follows, all of the results
are reported in terms of $Re(\nu)$ (energy) and $Im(\nu)$ (damping).

To solve Eq.~\ref{eq:isotropic}, a scalar and a 
vector potential are introduced which correspond
to dilatational and equivoluminal motions 
respectively.  Eq.~\ref{eq:isotropic} separates 
into scalar and vector parts.
Detailed expressions for the displacement
 $\vec u$ and the stress tensor 
$\tensor \sigma$ can be found in \cite{eringen}.
In a previous work \cite{PortalesPRB02}, we 
expressed $\vec u$ and 
$\tensor \sigma \cdot \vec r$ in
a more convenient basis.
Next, boundary conditions corresponding to the 
system have to be applied and result in secular 
equations whose complex valued roots gives the
frequencies of the eigenmodes of vibration.
Within the nanoparticle, the scalar and vector
potentials involve the spherical Bessel function
of the first kind $j_l$.

For method (A) there is no external force on
the sphere so that $\tensor \sigma \cdot \vec r = \vec 0$
at every point on the surface.
For method (B), the continuity of $\vec u$ and
$\tensor \sigma \cdot \vec r$ is imposed at the
surface of the nanoparticle where it contacts
the matrix.

If the glass matrix is a free sphere of radius $R_m$ 
concentric around the nanoparticle of radius $R_p$, normal mode
frequencies can be found \cite{PortalesPRB02}.  But what
happens is that every eigenfrequency vanishes as $R_{m}^{-1}$ as
$R_{m} \rightarrow \infty$. In other words, for a macroscopic
glass matrix there is a continuum of states but no discrete
modes.

The approach of method (B) is to look for
damped modes of vibration of the nanoparticle
where $\omega$ is now a complex number such that
$Re(\omega)$ is the frequency of the vibration
and $Im(\omega)$ is the damping. 
In the glass matrix, the spherical Hankel
function of the second kind, $h^{(2)}_{l}(k r)$,
is used because it corresponds to an outward
travelling wave. The wavevector $k$ is also complex.

Equations for the embedded sphere case of method (B)
have already been written by Dubrovskiy 
et al. \cite{dubrovskiy81}.
In order to find the roots of the secular
equation, Dubrovskiy et al. made
approximations which depend on the relative 
values of the parameters in the sphere and 
around it. We made no approximations
and found the complex roots numerically
\footnote{Programs are available upon request.}.
Frequencies and damping in \cite{Verma99} were reproduced with this
method.

Method (C) is a molecular dynamics calculation \cite{stillinger85}.
While this approach can readily be extended to the
more general case of an object of arbitrary shape
with inhomogeneous anisotropic elasticity, we choose for
brevity only to present the case of a spherical
object with cubic crystalline elasticity.
A simple cubic $a \times a \times a$ lattice of $N$
identical point particles of mass $\rho a^3$
interacting through "springs" has particle positions $\vec r_j$
integrated in time and Fourier transformed in order to
obtain mode frequencies. Each particle has six first
neighbors and twelve second neighbors.
Every pair $(i,j)$ of first neighbors
are coupled by potential energy 
$\frac{1}{2} k_{sp1} (\| \vec r_i - \vec r_j \|-a)^2$,
while second neighbor pairs $(i,j)$ are coupled by
$\frac{1}{2} k_{sp2} (\| \vec r_i - \vec r_j \|-\sqrt{2} a)^2$.
Cubic octets are coupled by
$ ( k_{8pt} / 4 a^2 ) (D - 6 a^2)^2$
where
$ D = \sum_{j=1}^{8} \| \vec R_{cm} - \vec r_j \|^2 $
and
$ \vec R_{cm} = \frac{1}{8} \sum_{j=1}^{8} \vec r_j $.
The bulk modulus is $ B = ( C_{11}+2 C_{12} ) / 3$. A small
isotropic dilatation $ \delta = dV/V $ stretches all neighbor
distances proportionally and leads to stored energy
$ \frac{1}{2} V B \delta^2 $, so that we obtain
$ a (C_{11} + 2 C_{12 } ) = k_{sp1} + 4 k_{sp2} + 24 k_{8pt} $.
Considering that a transverse plane wave moves along the
x-axis at speed $ \sqrt{C_{44} / \rho} $, we find
$ a C_{44} = k_{sp2} $.  Finally, considering that a longitudinal
plane wave moves along the x-axis at speed $ \sqrt{C_{11} / \rho} $,
we find
$ a C_{11} = k_{sp1} + 2 k_{sp2} + 8 k_{8pt} $.
Inversion allows the three force constants to be obtained for
a general cubically elastic material:
$k_{sp1} = a ( C_{11} - (C_{12} + C_{44} ) )$,
$k_{sp2} = a C_{44} $
and
$k_{8pt} = \frac{1}{8} a ( C_{12} - C_{44} )$.

The frequency dependence of a given mode is $f(N)$.
A sequence of simulations with
$N$ varying from a few hundred to 210000 is used to
extrapolate $f(\infty)$, which is the continuum limit.
Accurate results are obtainable with just a few thousand.
Actual nanoparticles are not continuum objects, 
and so the continuum limit is an idealization.  
Furthermore, our calculation uses the bulk 
elastic coefficients of silicon. Surface 
effects (arguably, the entire bulk of the nanoparticle
is close to its surface) and dispersion (effects of
the silicon crystal structure on phonon speed) are
not taken into account in this calculation.
However, the wavelengths of the lowest modes
are so large compared to interatomic distances that
dispersion will be negligible.

The correctness and degree of convergence
of method (C) was checked by comparing its results to
those of method (A) for isotropic elastic materials
of various Poisson ratios.
The results of the two methods agree closely.

Since the first report of low-frequency 
Raman scattering from nanoparticles \cite{Duval86},
such measurements have been made on many materials
\footnote{see \cite{Fujii96,PortalesPRB02,Ovsyuk96,Verma99} and
references therein.}.
Results are most often successfully 
explained with the free sphere model described above.
However, calculated frequencies are much too high
in silicon nanoparticles embedded in a
silica matrix \cite{Fujii96,pautheJMC99}.
In this work, we want to see if the situation 
can be improved by taking into account the 
presence of the matrix and the elastic 
anisotropy of silicon.

\section{Experiment}
\begin{figure}
\includegraphics[angle=270,width=\columnwidth]{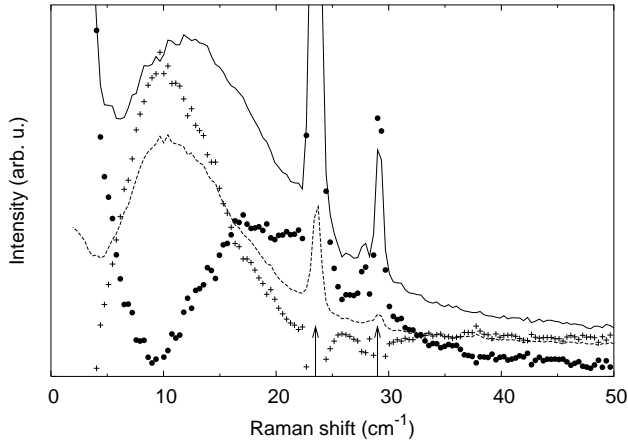}
\caption{\label{fig:raman}Parallel (solid line) and crossed
(broken line) configuration Raman spectra of 6.8 nm diameter
silicon nanoparticles embedded in silica.  A decomposition is
made into peaks with depolarization factor 0.75 (plus signs)
and 0.33 (bullets).  Laser plasma lines are indicated by
arrows.}
\end{figure}

New experimental results using the samples studied in \cite{pautheJMC99}
have been obtained. The experimental setup is a Jobin-Yvon T64000.
Incident light comes from the 514.5~nm line of an Argon ion laser
and is focused onto a 1~$\mu$m$^2$ area on the sample.  Power was kept
below 1~mW to avoid sample heating.  The CCD detector allows for long
accumulation times; the quality of the spectra is much improved compared
to previous measurements using a scanning monochromator coupled with
a photomultiplier.  Spectra obtained for the sample heat treated at
500~$^\circ$C are presented in Fig.~\ref{fig:raman}.  Measurements were
made with the sample at room temperature.  Average nanoparticle diameter
measured by transmission electron microscopy is 6.8 nm.
The two narrow lines centered around 23 and 29~cm$^{-1}$ are plasma
lines from the laser. The parallel and crossed spectra correspond
to scattered light intensity with polarization parallel
($I_{\parallel}$) and perpendicular ($I_{\perp}$)
to the excitation beam respectively.

The positions of the peak maxima of
these two spectra differ slightly indicating scattering by
several vibrational modes, one being strongly depolarized at
low frequency and another less depolarized contribution at higher
frequency.  From the ratio of the two spectra at 10~cm$^{-1}$
the depolarization factor of the first mode is close to
$\rho_{dep}$ = $I_{\perp}/I_{\parallel}=0.75$, as must be the
case for a non-resonant non-totally symmetric mode.
If we assume our spectra components to have only two different
depolarization factors, then the shape of the scattering with
depolarization factor different from 0.75 is given by
$I_{\parallel}*0.75-I_{\perp}$.
As a result, two vibrational modes are observed: the
first one with a maximum at 10~cm$^{-1}$ (determined from the
crossed configuration) and the second one with a broad maximum
around 20~cm$^{-1}$
(determined from the calculated spectra).

\section{Discussion}
First, we want to see how the free sphere 
model (A) compares with our
experimental results. 
Table~\ref{tab:free} presents the calculated
vibrational energies for a free isotropic 
sphere of diameter 6.8 nm
with the density of silicon and different 
sound speeds
corresponding to important crystal directions 
in silicon. The first and second $[110]$ lines correspond to 
transverse phonons polarized along
 $[\overline{1}10]$ and $[001]$ 
respectively. 
Uniformly averaging over all directions, the 
longitudinal and transverse speeds of sound in silicon
are (9017~$\pm$~233)~m/s and (5372~$\pm$~385)~m/s, respectively,
where one standard deviation of the variation is indicated.
The average rms speed is nearly the same.
For comparison, energies were also calculated using these
averaged sound speeds.
Elastic parameters were taken from \cite{HallPR67}.
As reported by Fujii et al. \cite{Fujii96}, these 
frequencies can't explain the lowest band position. 
Nevertheless, we will use these results to 
choose the isotropic parameters which best 
describe the calculated anisotropic energies.
 From Table~\ref{tab:free}, the closest agreement with
the exact anisotropic calculation (C) is obtained when
method (A) uses the averaged sound speeds.

\begin{table}
\caption{\label{tab:free}Energy ($\nu$ in 
cm$^{-1})$ of the fundamental spheroidal (SPH)
and torsional (TOR) vibration 
mode of a 6.8 nm diameter free silicon 
nanoparticle for different sound 
speeds using methods (A) and (C). $Im(\nu)$=0.}
\begin{ruledtabular}
  \begin{tabular}{ccddddd}
  \multicolumn{1}{c}{Calculation} & \multicolumn{1}{c}{Speed of} &
  \multicolumn{3}{c}{SPH} & \multicolumn{2}{c}{TOR}\\
method & sound & \multicolumn{1}{c}{$l$=0} & \multicolumn{1}{c}{$l$=1} & 
\multicolumn{1}{c}{$l$=2} & \multicolumn{1}{c}{$l$=1} & \multicolumn{1}{c}{$l$=2}\\
(A) & $[100]$     & 28.3 & 26.4 & 23.6 & 52.5 & 22.8 \\
(A) & $[111]$     & 38.7 & 28.0 & 21.1 & 45.8 & 19.9 \\
(A) & $[110]$     & 38.8 & 26.0 & 19.3 & 42.0 & 18.3 \\
(A) & $[110]$     & 33.8 & 29.1 & 23.7 & 52.5 & 22.8 \\
(A) & average     & 35.4 & 28.2 & 22.0 & 48.3 & 21.0 \\
(C) & anisotropic & 34.9 & 27.0 & 19.4 & 50.4 & 19.4 \\
  \end{tabular}
\end{ruledtabular}
\end{table}

We applied method (B), with averaged sound speeds,
to silicon nanoparticles embedded in silica glass.
Elastic parameters for silica were taken from \cite{Ovsyuk96}
and \cite{Sokolov92}.
Results are presented in 
Table~\ref{tab:embedded}.  Results using method (A)
(free sphere) are also shown, for which
case there is no damping.
Modes can be spheroidal (SPH) or torsional (TOR).
$l$ is the orbital quantum number and 
$n$ is the index of the solution.
Blank spaces in the free case indicate that there is 
no equivalent to the corresponding embedded mode.

The silica matrix has three 
effects: it shifts energies ($Re(\nu)$),
it creates damping ($Im(\nu)$) and it also introduces 
new modes. One of these new modes is (TOR,$l$=1,$n$=0).
In the absence
of the matrix this would correspond to rigid rotation.
The matrix imposes a restoring force 
changing the rotation into a libration
(rigid rotational oscillation).
Other new modes have more
complicated origins. Considering modes
existing in the free case, the more their 
energies are shifted, the greater the
damping. This is consistent with a stronger 
contribution of the matrix. 
The damping values given in 
Table~\ref{tab:embedded} correspond
to the Raman homogeneous half width at half maximum.
The observed Raman linewidth should be greater mainly because of the 
nanoparticle size distribution inside our samples.

For comparison purposes, calculated energies and damping of vibrations
of a 6.8 nm diameter spherical cavity in silica as well as
around an infinitely hard, heavy sphere are presented
in Table~\ref{tab:matrix}. In particular, the lowest
spheroidal modes in Table~\ref{tab:embedded} correspond
approximately to modes around a rigid, heavy sphere.

\begin{table}
\caption{\label{tab:embedded}Energy ($Re(\nu)$ in cm$^{-1}$) for free 
(method (A)) and silica embedded (method (B)) 6.8 nm diameter
silicon nanoparticles for various modes.
In the embedded case damping ($Im(\nu)$ in cm$^{-1}$) is the half width at half maximum.}

\begin{ruledtabular}
  \begin{tabular}{cccddd}
       &   &   & \multicolumn{1}{c}{free (A) } & \multicolumn{2}{c}{embedded (B)}\\
  mode & $l$ & $n$ & \multicolumn{1}{c}{energy} & \multicolumn{1}{c}{energy} &
  \multicolumn{1}{c}{damping}\\
  \hline
  \hline
  SPH     & 0 & 0 &  35.4 &  38.3 & 13.9\\
                 &   & 1 &  85.1 &  85.5 & 10.9\\
                 &   & 2 & 130.5 & 130.8 & 10.5\\
  \hline
                 & 1 & 0 &       &   6.3 &  9.9\\
                 &   & 1 &       &   9.3 & 16.2\\
                 &   & 2 &  28.2 &  32.7 & 11.4\\
  \hline
                 & 2 & 0 &       &  18.8 & 18.2\\
                 &   & 1 &  22.0 &  22.5 & 10.1\\
                 &   & 2 &  40.2 &  46.4 & 14.0\\
  \hline
  \hline
  TOR      & 1 & 0 &       & 11.8 & 11.0\\
                 &   & 1 &  48.3 & 48.4 &  6.8\\
                 &   & 2 &  76.2 & 76.3 &  6.7\\
  \hline
                 & 2 & 0 &       &  6.5 & 12.2\\
                 &   & 1 &  21.0 & 26.1 & 13.7\\
                 &   & 2 &  59.7 & 59.9 &  7.3\\
  \end{tabular}
\end{ruledtabular}

\end{table}

\begin{table}
\caption{\label{tab:matrix}Energy and damping of a silica matrix
surrounding a 6.8 nm diameter spherical cavity and an infinitely
rigid and heavy sphere. Only even parity modes are shown.}
\begin{ruledtabular}
  \begin{tabular}{cccdddd}
       & $ $ & $ $ & \multicolumn{2}{c}{cavity} & \multicolumn{2}{c}{rigid heavy sphere}\\
  mode & $l$ & $n$ & \multicolumn{1}{c}{energy} & \multicolumn{1}{c}{damping} & \multicolumn{1}{c}{energy} & \multicolumn{1}{c}{damping}\\
  \hline
  \hline
  SPH     & 0 & 0 &  9.1 &  7.4 &      &     \\
  \hline
                 & 2 & 0 &  6.1 &  4.6 & 15.2 & 14.5\\
                 &   & 1 &  7.8 & 20.3 &      &     \\
                 &   & 2 & 23.3 & 11.1 &      &     \\
  \hline
  \hline
  TOR      & 1 & 0 &  5.1 &  8.8 &      &     \\
  \end{tabular}
\end{ruledtabular}
\end{table}

 From Table~\ref{tab:embedded}, two modes are good
candidates for the lowest Raman peak: 
(SPH,$l$=1,$n$=1) and (TOR,$l$=1,$n$=0).
But SPH modes with odd $l$ and TOR modes with even $l$ have odd
parity and should be Raman inactive \cite{Duval92}.
The calculated damping for (TOR,$l$=1,$n$=0) is higher
than the Raman peak HWHM.
Moreover, size and shape distributions\cite{pautheJMC99} and
temperature broadening also increase the Raman linewidth.
A possible explanation is that in our
model, the nanoparticle-matrix contact is perfect and therefore
more energy is radiated into the matrix than in our sample.

The second mode is in agreement with the calculated (SPH,$l$=2,$n$=0,1)
modes. Following \cite{montagna95}, we assume the depolarization
factor of this peak to be $1/3$. Therefore, we are able to fully decompose
each spectra. The decomposition for the parallel configuration is shown in
Fig.~\ref{fig:raman}.

\section{Conclusion}
The special demands  
of silicon nanoparticles embedded in silica require
consideration of mode broadening by the matrix,
mode shifting by the matrix, new modes created by
the matrix and the anisotropy of the silicon.
A strong candidate as the dominant feature of
the Raman spectra is the libration mode (TOR,$l$=1,$n$=0)
which would be Raman inactive without either
anisotropy or nonspherical shape \cite{Duval92}.

The approach we are using can easily be extended
to handle other features of realistic nanoparticles,
such as non-spherical shape, surface relaxation
and details of the nanoparticle-matrix interface.
The extension to inhomogeneous onion-like
particles where the parameters have
radial dependence can also be made.

Exact incorporation of anisotropy is not difficult,
but relatively accurate estimates of frequency are
possible using an isotropic model of the elasticity
as long as direction averaged speeds of sound are used.
This works best for modes with $l \leq 1$, since they
are at most three-fold degenerate with isotropy
and are not split by cubic crystalline anisotropy.

For an isotropic material, the $l$=2 spheroidal and torsional modes
are 5-fold degenerate.  This degeneracy is broken by the
anisotropy, into what MD simulations suggest is a doublet.

Overall, this study has provided a clearer basis for
interpreting Raman experiments where the acoustic impedance
of the nanoparticle is close to that of the matrix.  Crystal
elasticity is anisotropic in any case.
However, without theoretical calculations of the absolute
cross sections of Raman scattering it is possible only to
make tentative identification of the vibrational modes that
appear in the low frequency Raman spectrum.
\bibliography{calcul}
\end{document}